\documentclass{article}
\usepackage[T1]{fontenc} 
\usepackage[utf8]{inputenc} 
\usepackage{ismir,amsmath,cite,url}
\usepackage{graphicx}
\usepackage{color}
\usepackage{lineno}
\usepackage{booktabs}
\usepackage{makecell}
\usepackage{amsfonts} 
\usepackage{multirow}

\title{CLaMP: Contrastive Language-Music Pre-training for Cross-Modal Symbolic Music Information Retrieval}

\multauthor
{Shangda Wu$^1$ \hspace{1cm} Dingyao Yu$^{2}$ \hspace{1cm} Xu Tan$^2$ \hspace{1cm} Maosong Sun$^{1,3}$}
{ $^1$ Central Conservatory of Music, China \hspace{1cm} $^2$ Microsoft Research Asia\hspace{1cm}\\
$^3$ Tsinghua University, China\\
{\tt\small shangda@mail.ccom.edu.cn, \{v-dingyaoyu, xuta\}@microsoft.com, sms@tsinghua.edu.cn} \\
{\small\url{https://ai-muzic.github.io/clamp}}
}
\def\authorname{S. Wu, D. Yu, X. Tan, and M. Sun}

\usepackage[bookmarks=false,pdfauthor={\authorname},pdfsubject={\papersubject},hidelinks]{hyperref}

\begin{document}

\maketitle
\begin{abstract}
We introduce \textbf{CLaMP}: \textbf{C}ontrastive \textbf{La}nguage-\textbf{M}usic \textbf{P}re-training, which learns cross-modal representations between natural language and symbolic music using a music encoder and a text encoder trained jointly with a contrastive loss. To pre-train CLaMP, we collected a large dataset of 1.4 million music-text pairs. It employed text dropout as a data augmentation technique and bar patching to efficiently represent music data which reduces sequence length to less than 10\%. In addition, we developed a masked music model pre-training objective to enhance the music encoder's comprehension of musical context and structure. CLaMP integrates textual information to enable semantic search and zero-shot classification for symbolic music, surpassing the capabilities of previous models. To support the evaluation of semantic search and music classification, we publicly release WikiMusicText (WikiMT), a dataset of 1010 lead sheets in ABC notation, each accompanied by a title, artist, genre, and description. In comparison to state-of-the-art models that require fine-tuning, zero-shot CLaMP demonstrated comparable or superior performance on score-oriented datasets. Our models and code are available at \url{https://github.com/microsoft/muzic/tree/main/clamp}.
\end{abstract}
\section{Introduction}\label{sec:introduction}
Symbolic Music Information Retrieval (MIR) is a field that deals with the automatic analysis and retrieval of music based on symbolic representations such as sheet music or MIDI files. Symbolic MIR has numerous practical applications, including music genre classification \cite{DBLP:conf/adbis/Karydis06,DBLP:conf/cimca/KofodA08}, automatic music transcription \cite{DBLP:conf/ismir/BelloMS00,DBLP:conf/ismir/Raphael02}, and music recommendation systems \cite{walshaw2014statistical}. However, traditional symbolic MIR approaches based on handcrafted features are often limited in their ability to capture the complex nature of music.

Deep learning has become increasingly popular in symbolic MIR \cite{DBLP:conf/ismir/HawthorneSSME21,DBLP:conf/iclr/GardnerSMHE22,DBLP:journals/taslp/KongLSWW21,DBLP:conf/acl/ZengTWJQL21} due to its ability to extract complex and abstract music features from large datasets. However, obtaining sufficient labelled data can be costly and time-consuming, as most labelled symbolic music datasets are small in size \cite{DBLP:conf/dlfm/FerraroL18, ferreira_ismir_2019, DBLP:journals/corr/abs-2107-05223}. To address this issue, semantic search and zero-shot classification techniques can be used to retrieve and label extensive unlabelled data. These techniques enable the search for music by a given open-domain query (e.g., "\textit{upbeat music with a fast tempo}"), or the automatic identification of music characteristics based on customized labels without the need for training data.

\begin{figure}[t]
    \vspace{-1em}
    \centering
		\begin{minipage}{\textwidth} 
            \includegraphics[width=8.25cm]{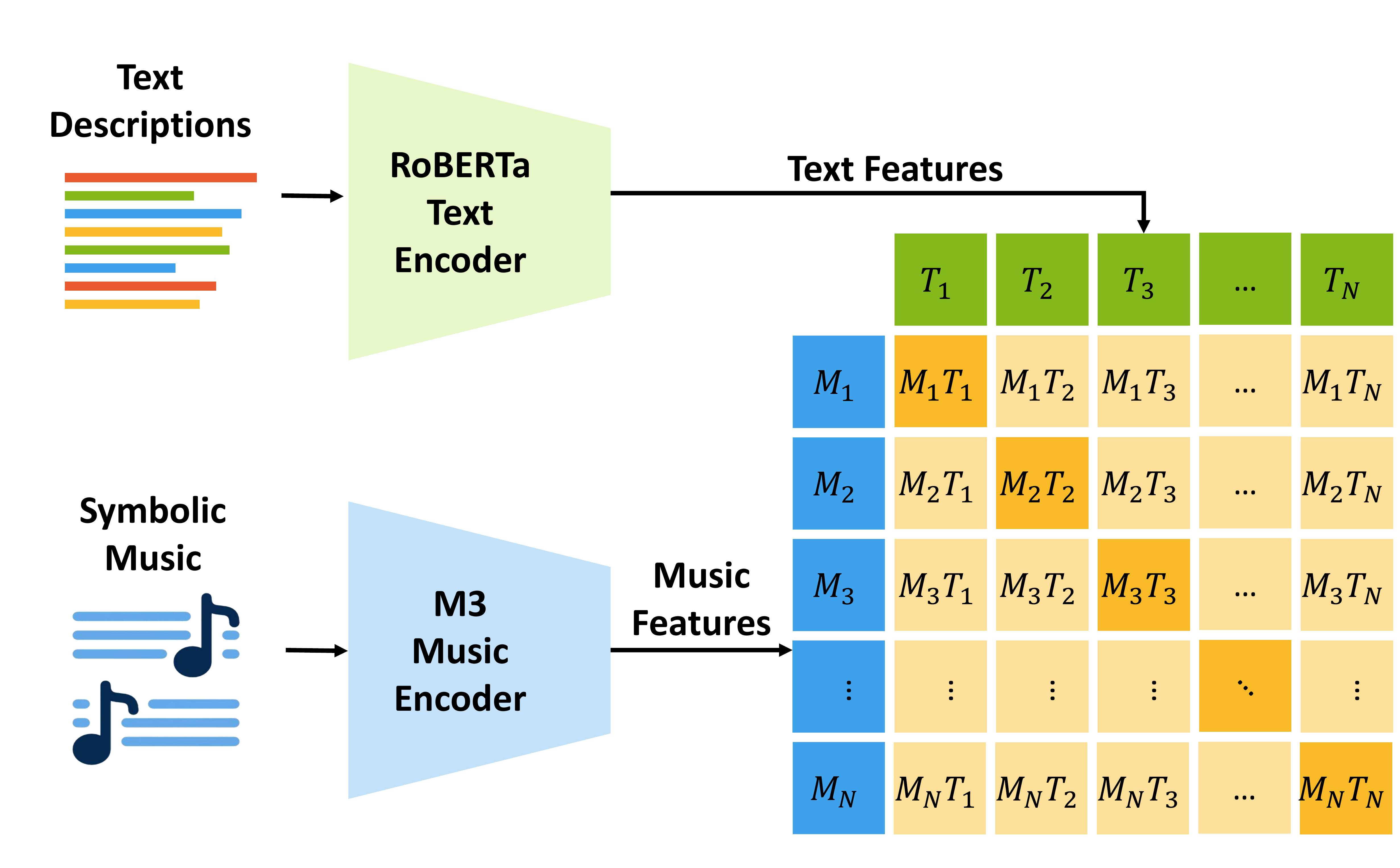}
		\end{minipage}
    \centering
	\caption{The architecture of CLaMP, including two encoders - one for music and one for text - trained jointly with a contrastive loss to learn cross-modal representations.}
\end{figure}

To enable semantic search and zero-shot classification for symbolic music, it is necessary to establish a connection between music and language. This can be achieved through the use of contrastive learning \cite{DBLP:conf/icml/RadfordKHRGASAM21,DBLP:conf/icml/KimSK21,DBLP:conf/icml/0001LXH22,DBLP:journals/corr/abs-2206-04769,DBLP:journals/corr/abs-2208-12415} and pre-training \cite{DBLP:conf/naacl/DevlinCLT19,DBLP:conf/nips/BrownMRSKDNSSAA20,DBLP:journals/jmlr/RaffelSRLNMZLL20}. Contrastive learning trains models to learn a feature space where similar sample pairs are grouped and dissimilar pairs are separated, while pre-training involves training a model on a large dataset that can be fine-tuned or directly applied to a specific task.

In this paper, we introduce a solution for cross-modal symbolic MIR that utilizes contrastive learning and pre-training. The proposed approach, \textbf{CLaMP}: \textbf{C}ontrastive \textbf{La}nguage-\textbf{M}usic \textbf{P}re-training, is inspired by the success of vision-language models \cite{DBLP:conf/icml/RadfordKHRGASAM21}. Unlike prior models that rely solely on symbolic music \cite{DBLP:conf/acl/ZengTWJQL21,DBLP:conf/ismir/0008X21,DBLP:journals/corr/abs-2107-05223}, CLaMP learns semantically rich representations of musical concepts from both sheet music and natural language. The contributions of this paper are as follows:

\begin{figure*}[t]
	\centering
		\begin{minipage}{\textwidth} 
            \includegraphics[width=\textwidth]{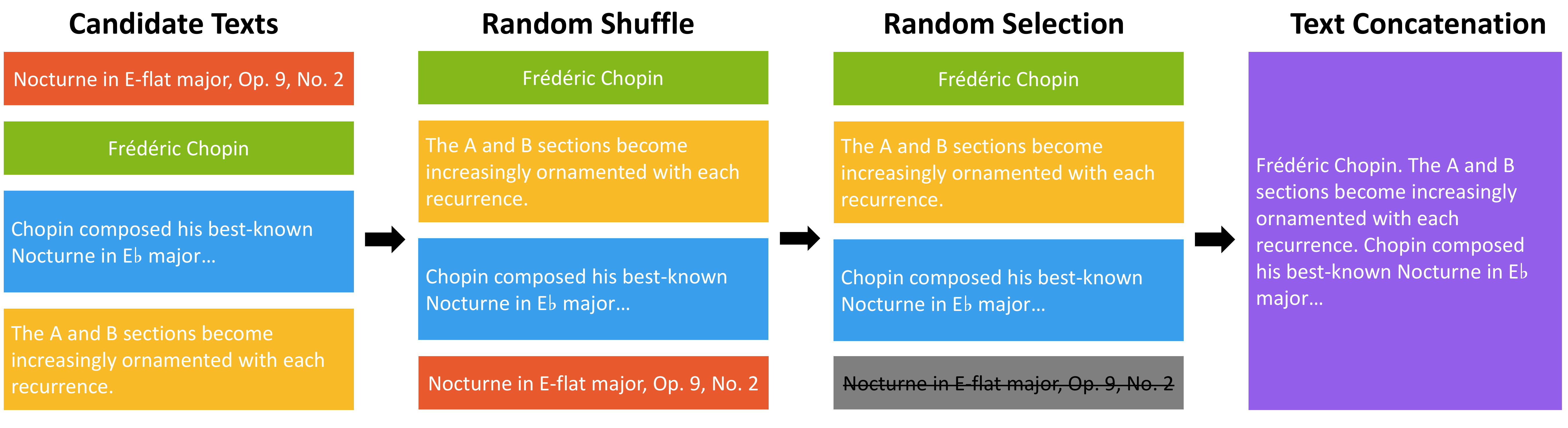}
		\end{minipage}
    \centering
	\caption{Text dropout is a data augmentation technique that involves a process in which candidate texts are shuffled randomly and then selected to form a concatenated text. In this example, three candidate texts were randomly selected and concatenated to produce the input text.} 
    \vspace{-1em}
\end{figure*}

\begin{itemize}
    \item
    CLaMP is a cross-modal model for symbolic MIR, which is pre-trained on WebMusicText (WebMT), a dataset of 1.4 million music-text pairs. To the best of our knowledge, this is the first model of its kind and it achieves comparable or better performance than existing state-of-the-art models without training.
    \item
    We propose multiple techniques to improve contrastive language-music pre-training. Our proposed techniques include applying text dropout as a data augmentation method, utilizing bar patching for efficient music representation, and implementing the masked music model pre-training objective.
    \item
    The cross-modal pre-training empowers CLaMP to perform tasks beyond the capabilities of unimodal models. It possesses unique features such as semantic search for desired music using open-domain text queries and zero-shot classification for new music.
    \item
    To facilitate the evaluation of semantic search and music classification, we release the WikiMusicText (WikiMT) dataset, which consists of 1010 music-text pairs sourced from Wikifonia and Wikipedia.
\end{itemize}

\section{Methodology}
This section presents CLaMP and its cross-modal symbolic MIR abilities. Additionally, we describe the WebMT dataset, which we created to pre-train our model.

\subsection{Model Design}
\subsubsection{Contrastive Learning Objective}
CLaMP jointly trains music and text encoders to represent the structural and semantic aspects of both modalities in a shared feature space. This is achieved using a batch construction method and objective \cite{DBLP:conf/nips/Sohn16,DBLP:journals/corr/abs-1807-03748}, as illustrated in Fig. 1, whereby the correct pairings of a batch of $N$ music-text pairs are predicted. The music and text encoders employ global average pooling to obtain corresponding features from the last hidden states.

The objective of CLaMP is to minimize the distance between $N$ paired music-text examples while maximizing the distance between $N^{2}-N$ unpaired examples. We denote a batch of $N$ music-text pairs as ${(m_i,t_i)}_{i=1}^N$, where $m_i$ and $t_i$ represent the $i$-th music and text inputs, respectively. The music and text encoders are represented as $f_m$ and $f_t$. The contrastive loss for ${(m_i,t_i)}_{i=1}^N$ is defined as follows:

\begin{footnotesize}
\begin{equation}
\begin{split}
\mathcal{L}_{CL} = -\frac{1}{2N}\sum_{i=1}^{N}( \log \frac{\exp(f_m(m_i)\cdot f_t(t_i)/\tau)}{\sum_{j=1}^{N} \mathbb{1}_{i\neq j} \exp(f_m(m_i)\cdot f_t(t_j)/\tau)} + \\
\log \frac{\exp(f_m(m_i)\cdot f_t(t_i)/\tau)}{\sum_{j=1}^{N} \mathbb{1}_{i\neq j} \exp(f_m(m_j)\cdot f_t(t_i)/\tau)}),
\end{split}
\end{equation}
\end{footnotesize}

\noindent
where $\tau$ is a temperature hyper-parameter that controls the sharpness of the softmax distribution, and $\mathbb{1}_{i\neq j}$ is an indicator function that equals 1 if $i\neq j$, and 0 otherwise. The two terms in Eq. 1 consider either music-to-text or text-to-music logits.

\subsubsection{Text Encoder}
CLaMP includes a text encoder to extract musically relevant features from the input text. To achieve optimal performance, a pre-trained language model is used to initialize the text encoder. Furthermore, text dropout is employed as a data augmentation technique to prevent overfitting and improve the generalization ability of the text encoder.

\textbf{Pre-trained Language Model} \quad RoBERTa \cite{DBLP:journals/corr/abs-1907-11692} is a transformer-based language model pre-trained on a large corpus of English text using the Masked Language Modeling (MLM) objective \cite{DBLP:conf/naacl/DevlinCLT19}. This model is designed to be fine-tuned on downstream tasks and has demonstrated excellent performance as a text encoder for the contrastive language-audio pre-training \cite{DBLP:journals/corr/abs-2211-06687}. To improve training efficiency, we used DistilRoBERTa \cite{Sanh2019DistilBERTAD} instead, which has fewer parameters (82M) compared to RoBERTa-base (125M) while achieving comparable performance.

\textbf{Text Dropout} \quad Text dropout is a data augmentation technique that encourages models to learn robust features from input texts. This technique involves using a dataset consisting of multiple paired candidate texts from various sources for each musical composition. Similar to \cite{10094670}, for a given composition with $L$ candidates, text dropout shuffles the set of candidate texts and randomly selects $K$ texts, where $K$ is uniformly and randomly sampled from integers ranging from 1 to $L$. These selected texts are concatenated to form a single input text for the text encoder, as shown in Fig. 2. Text dropout offers a wider range of possible text combinations and allows the model to learn more complex and diverse textual features.

\subsubsection{Music Encoder}
The CLaMP music encoder is designed to understand the complex musical structure and context within ABC notation. As a text-based format for symbolic music, ABC notation incorporates a wide range of musical symbols commonly used in sheet music. To keep all musical information while shortening sequence length, the encoding process utilizes the bar patching technique. To optimize performance, the music encoder is specifically designed for symbolic music understanding based on bar patching.

\begin{table}[t!]
    \vspace{-1em}
  \begin{center}
    \caption{The average number of tokens per lead sheet in the WikiMT dataset with different encoding methods.}

    \resizebox{\linewidth}{!}{\begin{tabular}{p{1.5cm}p{1.9cm}<{\centering}p{2.1cm}<{\centering}p{2.6cm}<{\centering}}
      \toprule 
      \textit{Encoding} & Bar Patching & ABC Notation & OctupleMIDI \cite{DBLP:conf/acl/ZengTWJQL21}\\
      \midrule 
      \textit{Tokens} & \textbf{47.07±21.60} & 749.16±379.56 & 469.09±256.43\\
      \bottomrule 
    \end{tabular}}
  \end{center}
  \vspace{-1em}
\end{table}

\textbf{Bar Patching} \quad The bar in musical notation groups phrases by defining a fixed number of beats and each bar can be read and played as a single unit. It is separated by vertical lines, providing reference points for locating positions within a score.

Previous models \cite{DBLP:journals/corr/SturmSBK16,Geerlings2020InteractingWG,DBLP:journals/corr/abs-2301-02884,wu2023exploring} for ABC notation utilized character-based tokenization, resulting in sequences that are too lengthy to process efficiently. On the other hand, MeasureVAE \cite{DBLP:conf/ismir/PatiLH19} demonstrated the feasibility of encoding scores at the bar-level for music generation. To improve the efficiency of processing, we proposed bar patching, inspired by patch-based techniques in computer vision \cite{DBLP:conf/iclr/DosovitskiyB0WZ21}.

Bar patching divides a score into several small segments corresponding to bars or headers (i.e. meta-information) in ABC notation. In our implementation, each patch is assigned a maximum of 64 characters, covering 98.8\% of the headers or bars in the pre-training dataset. We add an \texttt{[END]} token at the end of each patch to indicate the end of the sequence. Patches with fewer than 64 are padded with \texttt{[PAD]} tokens, while those with over 64 characters are truncated. For the vocabulary, 95 ASCII printable characters and three special tokens (i.e., \texttt{[PAD]}, \texttt{[MASK]}, and \texttt{[END]}) are considered, resulting in a total of 98 tokens. Thus, each patch can be represented as a 64$\times$98 matrix. These patches are then flattened and projected into 768 dimensions embeddings and used as input tokens, as illustrated in Fig. 3.

Bar patching effectively reduces the average sequence length of the encoded music to less than 10\% of the original ABC notation, as shown in Table 1. This technique improves the efficiency of representing music and facilitates faster computation while preserving all musical information in the notation.

\begin{figure}[t]
	\centering
		\begin{minipage}{\textwidth} 
            \includegraphics[width=8.25cm]{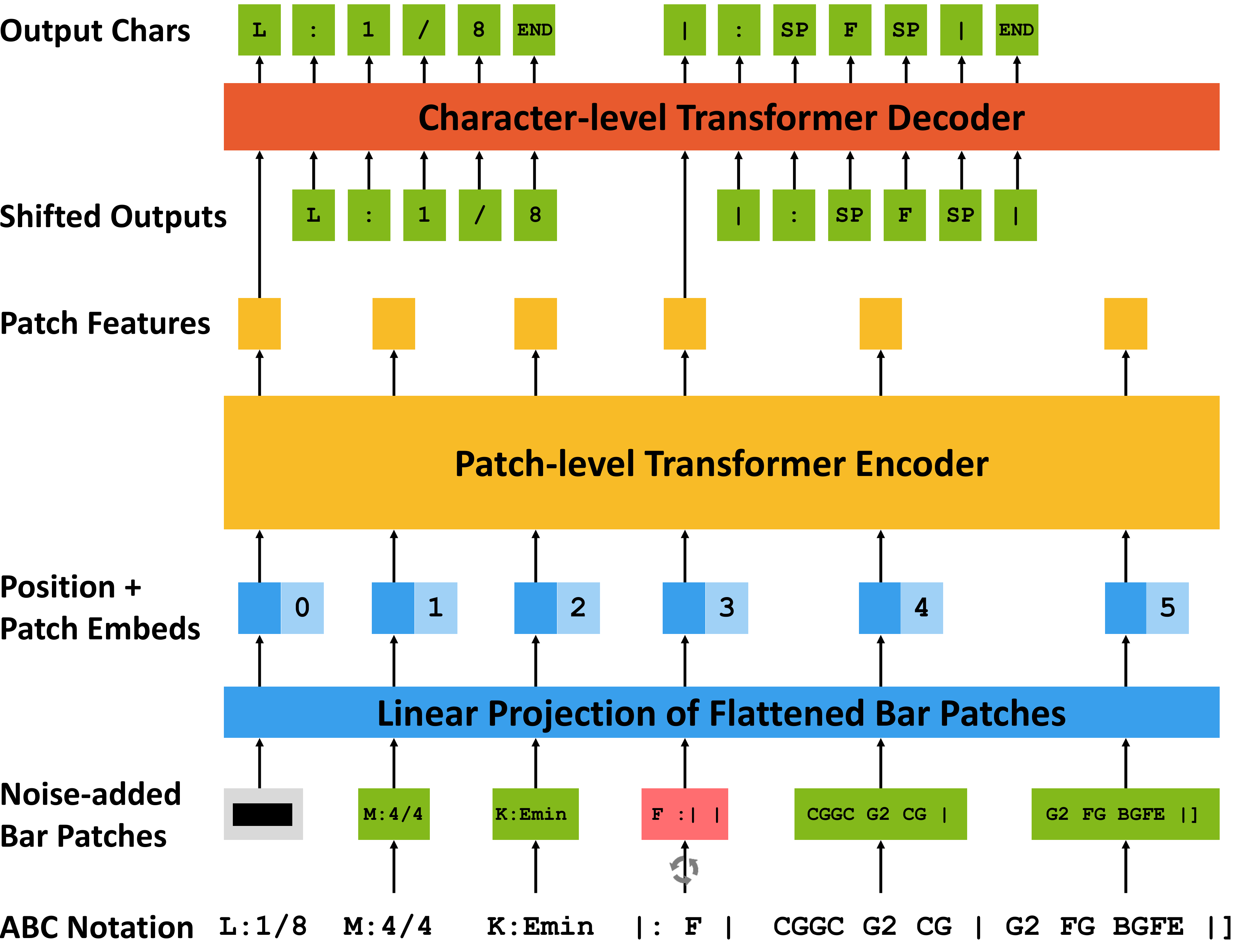}
		\end{minipage}
    \centering
	\caption{The masked music model architecture, where the encoder takes in a sequence of patches, and the decoder reconstructs character information of noise-added patches.}
\end{figure}

\begin{figure*}[t]
	\centering
		\begin{minipage}{\textwidth} 
            \includegraphics[width=\textwidth]{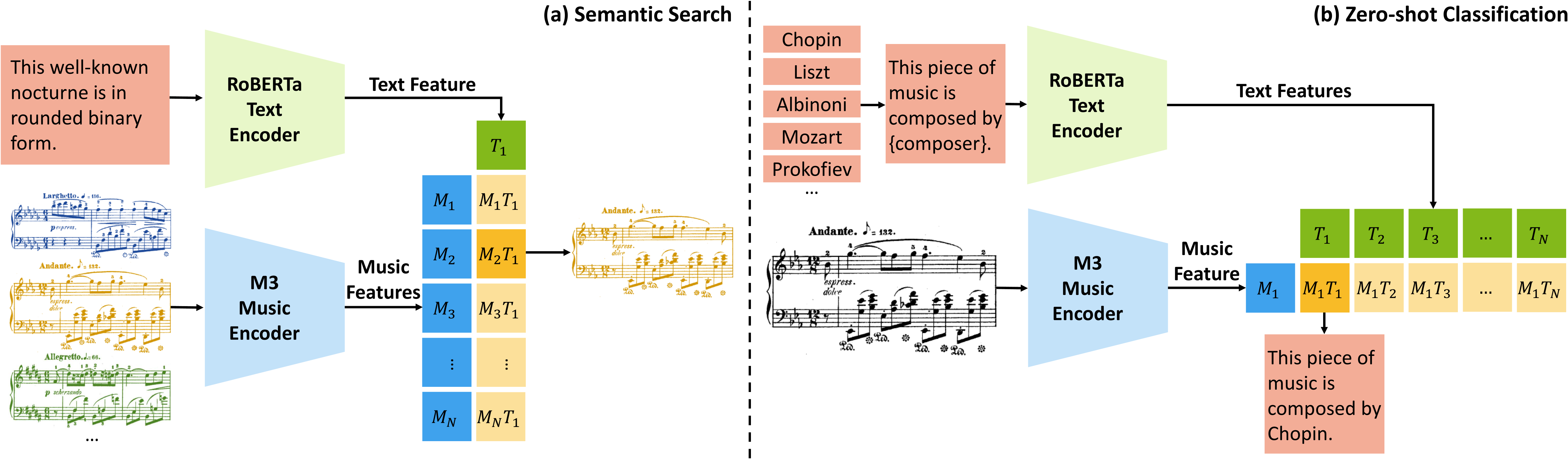}
		\end{minipage}
    \centering
    \caption{The processes of CLaMP performing cross-modal symbolic MIR tasks, including semantic search and zero-shot classification for symbolic music, without requiring task-specific training data.} 
    \vspace{-1em}
\end{figure*}

\textbf{Masked Music Model} \quad The Masked Music Model (M3) is a self-supervised model for symbolic MIR based on bar patching representation. The primary concept of M3 is to introduce random noise to certain patches of the input music, and then reconstruct the characters in the noise-added bar patches based on the context. This pre-training enables M3 to learn from unlabelled musical data, making it useful for initializing the CLaMP music encoder.

M3 is based on an asymmetric encoder-decoder architecture, similar to MAE \cite{DBLP:conf/cvpr/HeCXLDG22}, as shown in Fig. 3. It uses an encoder to extract contextualized features of individual patches, along with a decoder, which is lightweight and autoregressively reconstructs the characters for each patch. After pre-training, the decoder is discarded and the encoder is used to initialize the music encoder of CLaMP.

The pre-training objective is inspired by MLM \cite{DBLP:conf/naacl/DevlinCLT19}. We first randomly select $M\%$ of the bar patches in the input music, and then the noise is added in three different ways:

\begin{itemize}
    \item
    \textit{Masking:} $80\%$ of the selected bar patches are replaced with a special patch filled with \texttt{[MASK]} tokens. This encourages the model to learn to fill in missing information and understand the relationship between different musical elements.
    \item
    \textit{Shuffling:} $10\%$ of the selected bar patches are randomly shuffled internally. For example, a bar patch "\texttt{|: F |}" may be randomly shuffled to "\texttt{F :| |}" as shown in Fig. 3. This forces the model to learn the patterns and structures within bar patches.
    \item
    \textit{Unchanged:} $10\%$ of the selected bar patches are left unchanged. This can narrow down the gap between pre-training and fine-tuning.
\end{itemize}

M3 is trained to predict the original characters in the noise-added bar patches based on contextualized patch features. The model is optimized using the cross-entropy loss, which compares the predicted characters with the ground truth characters. The final objective is to minimize the average loss over all the noise-added bar patches in the training set. By denoising these bar patches, M3 learns to capture the dependencies and relationships between different musical elements and structures, allowing it to extract meaningful features from ABC notation.

\subsection{Cross-Modal Symbolic MIR}
CLaMP is capable of aligning symbolic music and natural language, which can be used for various cross-modal retrieval tasks, including semantic search and zero-shot classification for symbolic music. 

Semantic search is a technique for retrieving music by open-domain queries, which differs from traditional keyword-based searches that depend on exact matches or meta-information. This involves two steps: 1) extracting music features from all scores in the library, and 2) transforming the query into a text feature. By calculating the similarities between the text feature and the music features, it can efficiently locate the score that best matches the user's query in the library.

Zero-shot classification refers to the classification of new items into any desired label without the need for training data. It involves using a prompt template to provide context for the text encoder. For example, a prompt such as "\texttt{This piece of music is composed by \{composer\}.}" is utilized to form input texts based on the names of candidate composers. The text encoder then outputs text features based on these input texts. Meanwhile, the music encoder extracts the music feature from the unlabelled target symbolic music. By calculating the similarity between each candidate text feature and the target music feature, the label with the highest similarity is chosen as the predicted one.

\subsection{WebMusicText Dataset}
To facilitate the learning of relationships between natural language and symbolic music, we developed a dataset named WebMusicText (WebMT) by crawling an extensive collection of music-text pairs from the web. Our dataset comprises 1,448,750 pairs of music-text data, where all music files are in score-oriented formats (e.g., MusicXML, LilyPond, and ABC notation). To reduce the disparity between scores in different notations, we first converted all music files to MusicXML and then to ABC notation\footnote{\url{https://wim.vree.org/svgParse/xml2abc.html}}. In addition, to avoid information leakage, we removed any natural language (e.g., titles, composers, and lyrics) in ABC notation. The text parts of each pair were obtained from corresponding meta-information (e.g., title and composer) or user comments, and are all in English. WebMT features diverse musical compositions, from monophonic folk music to polyphonic orchestral music, which enables the model to learn a wide range of musical information.

\section{Experiments}
\subsection{Settings}
\subsubsection{Models}
\begin{itemize}
    \item \textbf{MusicBERT \cite{DBLP:conf/acl/ZengTWJQL21}}: This model combines unsupervised pre-training with supervised fine-tuning, which achieved state-of-the-art results. MusicBERT is available in two settings: \textit{\texttt{MusicBERT-S/1024}} ($\rm MusicBERT_{small}$), and \textit{\texttt{MusicBERT-B/1024}} ($\rm MusicBERT_{base}$). MusicBERT-S/1024 consists of 4 layers and was pre-trained on the small-scale Lakh MIDI Dataset (LMD, 148,403 pieces) \cite{raffel2016learning}, while MusicBERT-B/1024 has 12 layers and was pre-trained on the large-scale Million MIDI Dataset (MMD, 1,524,557 pieces). Both models have a maximum length of 1024.

    \item \textbf{M3}: Our proposed music encoder is used to compare the performances of unimodal and multimodal models trained on the same dataset (i.e., WebMT). M3 comes with two settings: \textit{\texttt{M3-S/512}} and \textit{\texttt{M3-S/1024}}, with maximum lengths of 512 and 1024, respectively. In the following experiments, both settings use the 6 encoder layers only.

    \item \textbf{CLaMP}: Several variants were tested to verify the effectiveness of the proposed techniques for improving contrastive language-music pre-training. These include \textit{\texttt{CLaMP-S/512}} which is the full model, \textit{\texttt{CLaMP-S/512 (w/o TD)}} which removes text dropout, \textit{\texttt{CLaMP-S/512 (w/o M3)}} which has a randomly initialized music encoder, and \textit{\texttt{CLaMP-S/512 (w/o M3, BP)}} which removes both M3 and bar patching, and uses char-level tokenization to encode raw ABC notation instead. \textit{\texttt{CLaMP-S/1024}} was included to verify the effectiveness of an extended maximum length.
\end{itemize}

\subsubsection{Pre-training}
The text encoder was initialized using DistilRoBERTa \cite{Sanh2019DistilBERTAD}, with a maximum length of 128, and the music encoder was initialized using two settings: M3-S/512 and M3-S/1024. A length of 512 resulted in truncating 17.29\% of compositions in WebMT, while a length of 1024 reduced truncation to 7.7\%. Both models were trained for 40 epochs with 6 encoder layers and 3 decoder layers, an embedding size of 768, and a noise ratio of 45\%. Based on these two M3 encoders, we developed CLaMP-S/512 and CLaMP-S/1024. Both of them were trained for 20 epochs, using the AdamW optimizer \cite{DBLP:conf/iclr/LoshchilovH19} with $\beta_{1}=0.9$, $\beta_{2}=0.999$, $\epsilon=10^{-8}$, and a weight decay coefficient of 0.01. The batch size is set to 640, and the temperature $\tau=0.2$. The training process was accelerated and memory was saved by using mixed precision \cite{DBLP:conf/iclr/MicikeviciusNAD18}.

\subsubsection{Evaluation Datasets}
We introduce WikiMusicText (WikiMT)\footnote{\url{https://huggingface.co/datasets/sander-wood/wikimt}}, a new dataset for the evaluation of semantic search and music classification. It includes 1010 lead sheets (melodies with harmonies) in ABC notation sourced from Wikifonia, each accompanied by a title, artist, genre, and description. The title and artist information is extracted from the score, whereas the genre labels are obtained by matching keywords from the Wikipedia entries and assigned to one of the 8 classes that loosely mimic the GTZAN genres \cite{DBLP:conf/mm/Sturm12}. The description is obtained by utilizing BART-large \cite{DBLP:conf/acl/LewisLGGMLSZ20} to summarize and clean the corresponding Wikipedia entry. Additionally, following WebMT, the natural language information within the ABC notation is removed.

In addition to WikiMT, we use two other datasets to evaluate music classification: VGMIDI and Pianist8. VGMIDI \cite{ferreira_ismir_2019} includes 204 score-oriented MIDI arrangements that were classified according to the valence-arousal model. Pianist8 \cite{DBLP:journals/corr/abs-2107-05223} contains symbolic piano performances of 411 pieces from 8 composers with distinct styles, which were automatically transcribed from audio using a model presented in \cite{DBLP:journals/taslp/KongLSWW21}.

\subsubsection{Metrics}
We use the following three metrics to evaluate the effectiveness of models in various downstream tasks: 

\begin{itemize}
\item Mean Reciprocal Rank (MRR) is used to evaluate ranking systems. This metric calculates the average of the reciprocal ranks of the correct answers, which measures the effectiveness of the ranking.

\item Hit Ratio at K (HR@K) measures the accuracy of the model by checking if the correct item is among the top K recommendations, which is often used in recommendation systems.

\item F1-macro score is a metric that assesses the overall effectiveness of a classification model. It is computed using the arithmetic mean (i.e., unweighted mean) of all the per-class F1 scores.
\end{itemize}

\begin{table}[t!]
  \begin{center}
    \caption{Semantic search performance of CLaMP on WikiMT (1010 music-text pairs) under different settings.}
    \resizebox{\linewidth}{!}{\begin{tabular}{p{2.8cm}p{1cm}<{\centering}p{1cm}<{\centering}p{1.1cm}<{\centering}p{1.2cm}<{\centering}}
      \toprule 
      \textit{Setting} & MRR & HR@1 & HR@10 & HR@100\\
      \midrule 
      \textit{S/512} & \textbf{0.2561} & \textbf{0.1931} & \textbf{0.3693} & \textbf{0.7020}\\
      \textit{S/1024} & 0.2016 & 0.1436 & 0.3109 & 0.6554\\
      \textit{S/512 (w/o TD)} & 0.1841 & 0.1248 & 0.2911 & 0.6188\\
      \textit{S/512 (w/o M3)} & 0.1262 & 0.0802 & 0.1960 & 0.5119 \\
      \textit{S/512 (w/o M3, BP)} & 0.0931 & 0.0525 & 0.1584 & 0.4426 \\
      \bottomrule 
    \end{tabular}}
  \end{center}
\end{table}

\begin{table*}[t!]
  \centering
  \caption{Classification performance of different models on three datasets: WikiMT (1010 pieces, 8 genres), VGMIDI (204 pieces, 4 emotions), and Pianist8 (411 pieces, 8 composers).}
  \vspace{1em}
  \scalebox{1}{\begin{tabular}{p{5cm}p{1.4cm}<{\centering}p{1.4cm}<{\centering}p{1.4cm}<{\centering}p{1.4cm}<{\centering}p{1.4cm}<{\centering}p{1.4cm}<{\centering}} 
    \toprule
    \multirow{2}{*}{\textit{Model}} & \multicolumn{2}{c}{WikiMT} & \multicolumn{2}{c}{VGMIDI \cite{ferreira_ismir_2019}} & \multicolumn{2}{c}{Pianist8 \cite{DBLP:journals/corr/abs-2107-05223}}\\
    \cmidrule{2-7}
    & \multicolumn{1}{c}{\textit{F1-macro}} & \multicolumn{1}{c}{\textit{Accuracy}} & \multicolumn{1}{c}{\textit{F1-macro}} & \multicolumn{1}{c}{\textit{Accuracy}} & \multicolumn{1}{c}{\textit{F1-macro}} & \multicolumn{1}{c}{\textit{Accuracy}}\\
    \midrule
    \textit{Linear Probe MusicBERT-S/1024} & 0.2401 & \textbf{0.3507} & 0.4662 & 0.5350 & 0.8047 & 0.8102\\
    \textit{Linear Probe MusicBERT-B/1024} & 0.1746 & 0.3219 & 0.5127 & 0.5850 & \textbf{0.8379} & \textbf{0.8413}\\
    \textit{Zero-shot CLaMP-S/512} & \textbf{0.2660} & 0.3248 & \textbf{0.5217} & \textbf{0.6176} & 0.2180 & 0.2512\\
    \textit{Zero-shot CLaMP-S/1024} & 0.2248 & 0.3406 & 0.4678 & 0.5049 & 0.1509 & 0.2390\\
    \midrule
    \midrule
    \textit{Linear Probe M3-S/512} & 0.2832 & 0.3990 & 0.5991 & 0.6667 & 0.6773 & 0.6909\\
    \textit{Linear Probe M3-S/1024} & 0.3079 & 0.4020 & 0.5966 & 0.6522 & 0.6844 & 0.6958\\
    \textit{Linear Probe CLaMP-S/512} & \textbf{0.3452} & 0.4267 & \textbf{0.6453} & \textbf{0.6866} & 0.7067 & 0.7152\\
    \textit{Linear Probe CLaMP-S/1024} & 0.3449 & \textbf{0.4416} & 0.6345 & 0.6720 & \textbf{0.7271} & \textbf{0.7298}\\
    \bottomrule
  \end{tabular}}
\end{table*}

\subsection{Results}
\subsubsection{Semantic Search}
In the semantic search evaluation, we assessed different versions of CLaMP for semantic search, aiming to test the efficacy of contrastive language-music pre-training techniques. The pre-training dataset WebMT and the evaluation dataset WikiMT have no overlap, thus guaranteeing the validity of our evaluation results. In addition, as semantic search requires no additional training for this dataset, it demonstrates the generalizability of CLaMP. 

Table 2 shows that our full model (CLaMP-S/512) outperforms all other models across all metrics. Interestingly, we discovered that increasing the maximum sequence length to 1024 (CLaMP-S/1024) did not lead to an improvement in performance. We attribute this to the fact that all lead sheets in the WikiMT dataset, once encoded with bar patching, have a length smaller than 512, which limits the potential advantages of the longer sequence length of CLaMP-S/1024. We also observed that the removal of the proposed techniques from CLaMP had a considerable negative impact on semantic search performance. Notably, the removal of M3 pre-training had the greatest effect on model performance, followed by text dropout and bar patching.

In conclusion, our evaluation of CLaMP on WikiMT shows that CLaMP-S/512 with all proposed contrastive language-music pre-training techniques is the most effective for the semantic search task. This highlights the importance of these techniques for effective pre-training and semantic search tasks. Additionally, increasing the sequence length (CLaMP-S/1024) did not improve the model's performance. These results emphasize the significance of using appropriate pre-training techniques in language-music models and suggest that a longer sequence length may not necessarily result in better outcomes.

\subsubsection{Music Classification}
The goal of the classification evaluation is to assess how well the zero-shot CLaMP models perform compared to other fine-tuned models. In addition, to evaluate pre-trained models, linear probes are used to train a linear classifier for the classification based on the features from pre-trained models. Despite being less powerful and relying on pre-trained model features, linear classifiers offer a valuable means of quantitatively assessing feature quality \cite{DBLP:conf/iclr/AsanoRV20}.

WikiMT was converted into the MIDI format using music21 \cite{DBLP:conf/ismir/CuthbertA10} to be compatible with MusicBERT. In contrast, for VGMIDI and Pianist8, we employed MuseScore3's batch conversion tool\footnote{\url{https://musescore.org/en/project/batch-convert}} to convert the scores into the MusicXML format, which were then converted into ABC notation for use with M3 and CLaMP.

We conducted 5-fold cross-validation with the same folds to assess all linear probe models, using identical fine-tuning settings and a batch size of 10 to ensure consistency, given the limited size of the evaluation datasets. The linear probe CLaMP models used the music encoder only, while the text encoder was discarded. In the zero-shot classification setting, CLaMP had no previous exposure to these evaluation datasets during pre-training. We utilized manually designed prompts for the zero-shot CLaMP models.

The top half of Table 3 presents the comparison of the performance between linear probe MusicBERT and zero-shot CLaMP. The results found that the zero-shot CLaMP models demonstrated comparable or even superior performance compared to the linear probe MusicBERT models on WikiMT and VGMIDI datasets. Interestingly, the smaller zero-shot CLaMP-S/512 outperformed the larger linear probe MusicBERT-B/1024, indicating that the pre-training of CLaMP has enabled it to learn more generalizable features that are useful for zero-shot music classification. However, this trend was not observed on Pianist8, where MusicBERT models performed much better than zero-shot CLaMP models. This difference in performance can be attributed to the source of the datasets, as WikiMT and VGMIDI primarily focus on score information, whereas Pianist8 contains performance MIDI data derived from audio. Since both CLaMP and M3 were trained exclusively on score information, they lack knowledge of performance MIDI. However, we noticed that the performances of linear probe CLaMP models on Pianist8 significantly improved after fine-tuning compared to the zero-shot ones. This suggests that incorporating ABC notation from performance MIDI into the pre-training of CLaMP may enhance its ability to comprehend such data.

The linear probe CLaMP models show better performance compared to the linear probe M3 models, as indicated in the bottom half of Table 3, despite being pre-trained on the same dataset with the same architecture. This is attributed to the use of contrastive learning, which aligns the music encoder of CLaMP with the text modality, thus implicitly introducing textual information to the music encoder. Furthermore, we found that CLaMP-S/1024 performed better on Pianist8 than CLaMP-S/512, suggesting that a larger maximum length is beneficial for models to learn performance MIDI.

In summary, our evaluation demonstrates that zero-shot CLaMP performs comparably to state-of-the-art models in music classification. Furthermore, the incorporation of contrastive learning and textual information enhances the music encoder's performance, resulting in better classification accuracy when compared to M3 which employed the same architecture. These results highlight the potential of CLaMP as a pre-training framework for symbolic MIR.

\section{Conclusions}
This paper introduces CLaMP, a pre-trained model that utilizes contrastive language-music pre-training techniques to build cross-modal representations between natural language and symbolic music. The model was trained on a dataset containing 1.4 million music-text pairs and has demonstrated unique abilities of semantic search and zero-shot classification for symbolic music. Compared to state-of-the-art models that require fine-tuning, zero-shot CLaMP exhibits comparable or superior performance in score-oriented music classification tasks without any training. However, the current version of CLaMP has limited comprehension of performance MIDI, and still has room for improvement. Future research will aim to expand its capabilities by scaling it up and pre-training it on larger datasets that incorporate a wider range of symbolic music formats beyond score-oriented ones. We expect that its cross-modal representations will facilitate research on new topics in music analysis, retrieval, and generation, and provide a foundation for the development of innovative systems and applications that integrate music and language.

\bibliography{ISMIRtemplate}

\newpage
\appendix

\begin{table}[t!]
  \begin{center}
    \caption{The genre keywords we used for automatically assigning genre labels based on  Wikipedia entries.}

    \resizebox{\linewidth}{!}{\begin{tabular}{p{1cm}p{7.5cm}<{\centering}}
      \toprule 
      \textit{Genre} & Keywords\\
      \midrule 
      \textit{Jazz} & jazz/bebop/swing\\
      \textit{Country} & country/bluegrass/honky-tonk\\
      \textit{Folk} & folk/singer-songwriter/acoustic\\
      \textit{R\&B} & r\&b/motown/funk/soul/gospel/blues/boogie-woogie\\
      \textit{Pop} & pop/bubblegum\\
      \textit{Rock} & rock/punk\\
      \textit{Dance} & dance/disco/electronic\\
      \textit{Latin} & latin\\
      \bottomrule 
    \end{tabular}}
  \end{center}
  \vspace{-1em}
\end{table}

\begin{table}[t!]
  \begin{center}
    \caption{Token count statistics for the WikiMT dataset with different encoding methods and data types.}

    \resizebox{\linewidth}{!}{\begin{tabular}{p{3.1cm}p{2cm}<{\centering}p{1cm}<{\centering}p{1cm}<{\centering}}
      \toprule 
      \textit{Encoding} & Avg & Max & Min\\
      \midrule 
      \textit{ABC notation (music)} & 749.16±379.56 & 2912 & 107\\
      \textit{Bar Patching (music)} & 47.07±21.60 & 188 & 11\\
      \textit{Whitespace (text)} & 54.63±13.94 & 99 & 22\\
      \textit{BPE (text)} & 72.96±17.60 & 127 & 31\\
      \bottomrule 
    \end{tabular}}
  \end{center}
  \vspace{-1em}
\end{table}

\begin{figure}[t]
	\centering
		\begin{minipage}{\textwidth} 
            \includegraphics[width=8.25cm]{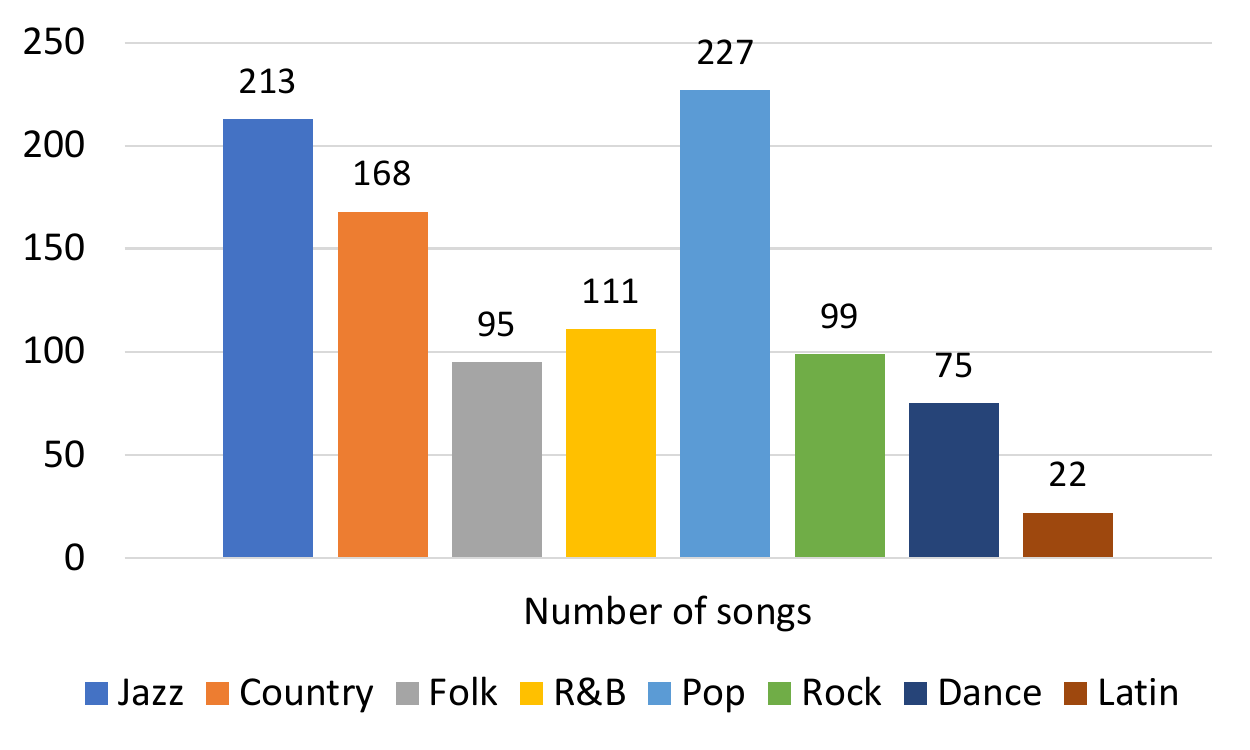}
		\end{minipage}
    \centering
	\caption{The genre distribution of the WikiMT dataset.}
\end{figure}

\section{WikiMusicText dataset}
\subsection{Data Collection and Curation}
To collect the data for the WikiMusicText (WikiMT) dataset, the following steps were taken: 1) converting all the MusicXML sheet music in Wikifonia to ABC notation, 2) extracting the title and artist name from each piece of music, 3) crawling the corresponding Wikipedia entry based on the title and artist name, 4) checking the most frequent genre keywords in the entry (as given in Table 4), and assigning the corresponding genre label, and 5) using BART-large to summarize the entry to keep them concise while also in a good form.

To be included in the WikiMT dataset, each piece of music had to meet the following criteria: 1) it must have a title and artist information, 2) the original entry must contain genre keywords, and only one genre could have the most matches, 3) after summarization by BART-large, the entry must still include the artist name, and 4) to ensure that there were sufficient data for each genre, any genres that had less than 10 files were removed from the dataset (e.g., classical, hip-hop, metal, and reggae).

All of the scores included in WikiMT were believed to be in the public domain to the best of our knowledge. However, it is important to acknowledge that copyright issues cannot be entirely ruled out. Therefore, users of the dataset should exercise caution when using it.

\subsection{Dataset Statistics}
WikiMT comprises 1010 pieces of music in ABC notation converted from Wikifonia. Each composition is accompanied by its title, artist, genre, and a summarized Wikipedia entry. The dataset covers eight genres, and their respective piece counts are displayed in Fig. 5.

Table 5 shows that the token counts for the dataset vary significantly depending on the encoding method and data type. ABC notation yields the highest token count due to its detailed and complex representation of sheet music. In contrast, the proposed bar patching representation reduces the length of music to less than 10\%. However, as all scores in WikiMT are lead sheets, the average sequence length of ABC notation is still acceptable for most models.

In terms of text encoding, whitespace encoding has a lower average token count than Byte-Pair Encoding (BPE). This is to be expected as whitespace encoding simply breaks text into tokens based on whitespace characters in the text, whereas BPE is subword tokenization, which decomposes rare words into smaller subwords.

WikiMT is a unique resource to support the evaluation of semantic search and music classification. However, it is important to acknowledge that the dataset was curated from publicly available sources, and there may be limitations concerning the accuracy and completeness of the genre and description information. Further research is needed to explore the potential biases and limitations of the dataset and to develop strategies to address them.

\section{Prompts for Music Classification}

\subsection{WikiMT}
\begin{itemize}
\item \textit{Country}: This piece of music is characterized by its roots in the traditional music of the American South, with its distinctive twangy sound and storytelling lyrics that often touch on themes of love, heartbreak, and rural life.
\item \textit{Folk}: This piece of music draws on the traditional music of a particular culture or region, often featuring acoustic instruments and simple, catchy melodies that are easy to sing along to. Folk music often tells stories of everyday life and the struggles of ordinary people.
\item \textit{Dance}: Whether it's disco, hip-hop, or EDM, dance music is all about creating a high-energy atmosphere and bringing people together through the universal language of dance.
\item \textit{Latin}: This piece of music draws on the rich musical traditions of Latin America, with its vibrant rhythms, colorful melodies, and passionate lyrics that often touch on themes of love, passion, and cultural identity.
\item \textit{Jazz}: This piece of music is characterized by its improvisational nature, with musicians often taking turns soloing over a complex and syncopated rhythm section. Jazz music often draws on elements of blues, swing, and Bebop music, and is known for its sophisticated harmonies and inventive melodies.
\item \textit{Pop}: This piece of music is designed to be catchy and easy to sing along to, with simple, memorable melodies and lyrics that often touch on themes of love, relationships, and self-expression. Pop music can take many forms, from bubblegum pop and boy bands to synthpop and EDM.
\item \textit{Rock}: Whether it's classic rock, punk, or grunge, rock music is all about pushing boundaries and challenging the status quo. Rock music has a rich history that spans decades, with iconic bands and legendary performances that continue to inspire new generations of musicians and fans.
\item \textit{R\&B}: R\&B music can take many forms, from classic Motown to contemporary hip-hop and trap soul.
\end{itemize}

\subsection{VGMIDI}
\begin{itemize}
    \item \textit{Joy (high valence high arousal)}: This piece of music is a jubilant celebration that radiates a contagious sense of happiness and joy.
    \item \textit{Anger (low valence high arousal)}: This piece of music is a visceral experience that unleashes a torrent of anger and fear.
    \item \textit{Sadness (low valence low arousal)}: This piece of music is a poignant reflection that evokes a deep sense of sadness and melancholy, taking the listener on an emotional journey through the depths of human experience.
    \item \textit{Calmness (high valence low arousal)}: This piece of music is a soothing balm that washes over the listener with a gentle wave of calmness and tranquility.
\end{itemize}

\subsection{Pianist8}
\begin{itemize}
\item \textit{Clayderman}: This music was performed by Richard Clayderman, a French pianist known for his romantic and sentimental style, whose repertoire includes a mix of original compositions, classical pieces, and popular music covers.
\item \textit{Yiruma}: This music was composed by Yiruma, a South Korean pianist and composer whose music has gained popularity through YouTube and social media, and whose style combines classical and contemporary elements with a strong emotional core.
\item \textit{Hancock}: This music was composed by Herbie Hancock, an American jazz pianist and composer who is known for his innovative approach to jazz music, and for incorporating elements of funk, rock, and electronic music into his compositions.
\item \textit{Einaudi}: This music was composed by Ludovico Einaudi, an Italian pianist and composer who is known for his minimalist and meditative music that often incorporates elements of classical, rock, and electronic music.
\item \textit{Hisaishi}: This music was composed by Hisaishi Joe, a Japanese composer known for his emotionally deep music that often features a blend of classical and traditional Japanese elements, and who has worked on numerous films and anime soundtracks.
\item \textit{Ryuichi}: This music was composed by Ryuichi Sakamoto, a Japanese musician and composer known for his eclectic and experimental approach to music, which often blends elements of classical, electronic, and traditional Japanese music, and who has worked on a variety of projects including film scores, solo albums, and collaborations with other musicians.
\item \textit{Bethel}: This music was created by the band Bethel Music, an American worship music collective based in California, known for its contemporary sound and strong Christian themes.
\item \textit{Hillsong}: This music was created by Hillsong, an Australian-based worship music collective that has become one of the most well-known and influential Christian music groups in the world, characterized by its powerful lyrics, modern sound, and uplifting messages of faith and hope.
\end{itemize}

%
%
%
%
%

\end{document}